\documentclass{segabs}
\usepackage[]{standalone}  % Use the standalone package
\usepackage{enumitem}  
\usepackage{url}
\usepackage{hyperref} 
\usepackage{import}
\usepackage{amsfonts}
\usepackage{lipsum}
\usepackage{float}
\usepackage{amsmath}
\usepackage{relsize}
\usepackage{float}
\usepackage{pgfplots} 
\usepackage{graphicx}
\DeclareMathAlphabet{\mathcal}{OMS}{cmsy}{m}{n}

 \pgfplotsset{compat=1.14}

\begin{document}
%\includepdf[]{coverpage}
%\input{template}

\onecolumn

\date{}% Certain latex templates automatically add the compilation date as a footnote in the generated pdf. This command can remove the date in some of the templates but does not work for all.

\onecolumn % make sure you keep this coverpage as one column. In this template, we force the coverpage to be one column with this command and then switch to double column for the remaining of the paper with the \doublecolumn command. 

%\begin{description}[labelindent=-3cm,leftmargin=1cm,style=multiline]
\begin{itemize}
\item[\textbf{Citation}]{A. Mustafa, M. Alfarraj, and G. AlRegib, “Estimation of Acoustic Impedance from Seismic Data using Temporal Convolutional Network,” Expanded Abstracts of the SEG Annual Meeting , San Antonio, TX, Sep. 15-20, 2019.}

%\item[\textbf{DOI}]{\url{https://doi.org/10.1109/MSP.2017.2783449}}

\item[\textbf{Review}]{Date of presentation: 18 Sep 2019}

\item[\textbf{Data and Code}]{\href{https://github.com/olivesgatech/Estimation-of-acoustic-impedance-from-seismic-data-using-temporal-convolutional-network}{[\underline{Github Link}]}}

\item[\textbf{Bib}] {@incollection\{amustafa2019AI,\\
title=Estimation of Acoustic Impedance from Seismic Data using Temporal Convolutional Network, \\
author=Mustafa, Ahmad and AlRegib, Ghassan, \\
booktitle=SEG Technical Program Expanded Abstracts 2019, \\
year=2019, \\
publisher=Society of Exploration Geophysicists\}
}

% Preprint sharing policy can vary depending on the publisher. Before posting a paper to arXiv, please specifically check the transaction/convference you are targeting. In some transactions, papers are usually added to arxiv after acceptance. Pubslishers usually allow the authors to share accepted version of their papers but not the final formatted version that is provided by the pubisher. In case of sharing preprints, publishers usually ask to add DOI and citation to the paper along with a copyright notice.

\item[\textbf{Contact}]{\href{mailto:amustafa9@gatech.edu}{amustafa9@gatech.edu}  OR \href{mailto:alregib@gatech.edu}{alregib@gatech.edu}\\ \url{http://ghassanalregib.com/} \\ }
\end{itemize}

%Following command sequence was used to start the paper content from the following page and avoid numbering cover page.
\thispagestyle{empty}
\newpage
\clearpage
\setcounter{page}{1}

%Cover page was 1 column. \twocolumn changes the page format back to double column.
\twocolumn

\title{Estimation of Acoustic Impedance from Seismic Data using Temporal Convolutional Network}
\renewcommand{\thefootnote}{\fnsymbol{footnote}} 
\author{Ahmad Mustafa\footnotemark[1], Motaz Alfarraj, and Ghassan AlRegib\\ Center for Energy and Geo Processing (CeGP), Georgia Institute of Technology}

\maketitle

\begin{abstract}
In exploration seismology, seismic inversion refers to the process of inferring physical properties of the subsurface from seismic data. Knowledge of physical properties can prove helpful in identifying key structures in the subsurface for hydrocarbon exploration. In this work, we propose a workflow for predicting acoustic impedance (AI) from seismic data using a network architecture based on Temporal Convolutional Network by posing the problem as that of sequence modeling. The proposed workflow overcomes some of the problems that other network architectures usually face, like gradient vanishing in Recurrent Neural Networks, or overfitting in Convolutional Neural Networks. The proposed workflow was used to predict AI on Marmousi 2 dataset with an average $r^{2}$ coefficient of $91\%$ on a hold-out validation set. 
\end{abstract}

\section{Introduction}
Reservoir characterization workflow involves the estimation of physical properties of the subsurface, like acoustic impedance (AI), from seismic data by incorporating knowledge of the well-logs. However, this is an extremely challenging task because in most seismic surveys due to the non-linearity of the mapping from seismic data to rock properties. Attempts to estimate physical properties from seismic data have been done using supervised machine learning algorithms, where the network is trained on pairs of seismic traces and their corresponding physical property traces from well-logs. The trained network is then used to obtain a map of physical properties for the entire seismic volume. 

Recently, there has been a lot of work integrating machine learning algorithms in the seismic domain \cite[]{regib}. The literature shows successful applications of supervised machine learning algorithms to estimate petrophysical properties. For examples, \cite[]{article} used Artificial Neural Networks to predict velocity from prestack seismic gathers, \cite[]{ALANAZI201264} used Support Vector Regression to predict porosity and permeability from core- and well-logs,  \cite[]{Chaki} proposed novel preprocessing schemes based on algorithms like Fourier Transforms and Wavelet Decomposition before using the seismic attribute data to predict well-log properties. More recently, \cite[]{GAN} used Generative Adversarial Networks (GANs) to map migrated seismic sections to their corresponding reflectivity section.
\cite[]{biswas} used Recurrent neural networks to predict stacking velocity from seismic offset gathers. \cite[]{alfarraj2018petrophysical} used Recurrent Neural Networks to invert seismic data for petrophysical properties by modeling seismic traces and well-logs as sequences. \cite[]{das2018} used Convolutional Neural Network (CNNs) to predict p-impedance from normal incident seismic. 

One challenge in all supervised learning schemes is to use a network that can train well on a limited amount of training data and can also generalize beyond the training data. Recurrent Neural Networks (RNNs) can subvert this problem by sharing their parameters across all time steps, and by using their hidden state to capture long term dependencies. However, they can be difficult to train because of the exploding/vanishing gradient problem. CNNs have great utility in capturing local trends in sequences, but in order to be able to capture long term dependencies, they need to have more layers (i.e, deeper networks), which in turn increase the number of learnable parameters. A network with a large number of parameters cannot be trained on limited training examples. 

In this work, we used Temporal Convolutional Networks (TCN) to modeling traces as sequential data. The proposed network is trained in a supervised learning scheme on seismic data and their corresponding rock property traces (from well logs). The proposed workflow encapsulates the best features of both RNNs and CNNs as is captures long term trends in the data without requiring a large number of learnable parameters.

\section{Temporal Convolutional Networks}
One kind of sequence modeling task is to map a given a sequence of inputs $\{x(0),...,x(T-1)\}$ to a sequence of outputs ${y(0),...,y(T-1)}$ of the same length, where $T$ is the total number of time steps. The core idea is that this kind of a mapping described by the equation \ref{eq:seq} can be represented by a neural network parameterized by $\Theta$ (i.e., $\mathcal{F}_{\Theta}$).

\begin{equation}
    \hat{y}(t) = \mathcal{F}\left(x(0),\dots,x(t)\right) \forall t\in [0,T-1]
    \label{eq:seq}
\end{equation}

Convolutional Neural Networks (CNNs) have been used extensively for sequence modeling tasks like document classification \cite[]{johnson}, machine translation \cite[]{Kalchbrenner}, audio synthesis\cite[]{Oord}, and language modeling\cite[]{Dauphin}. More recently, \cite[]{Bai} performed a thorough comparison of canonical RNN architectures with their simple CNN architecture, which they call the Temporal Convolutional Network (TCN), and showed that the TCN was able to convincingly outperform RNNs on various sequence modeling tasks.

\begin{figure}[ht]
\centering
  \includegraphics[width=0.5\linewidth
  ]{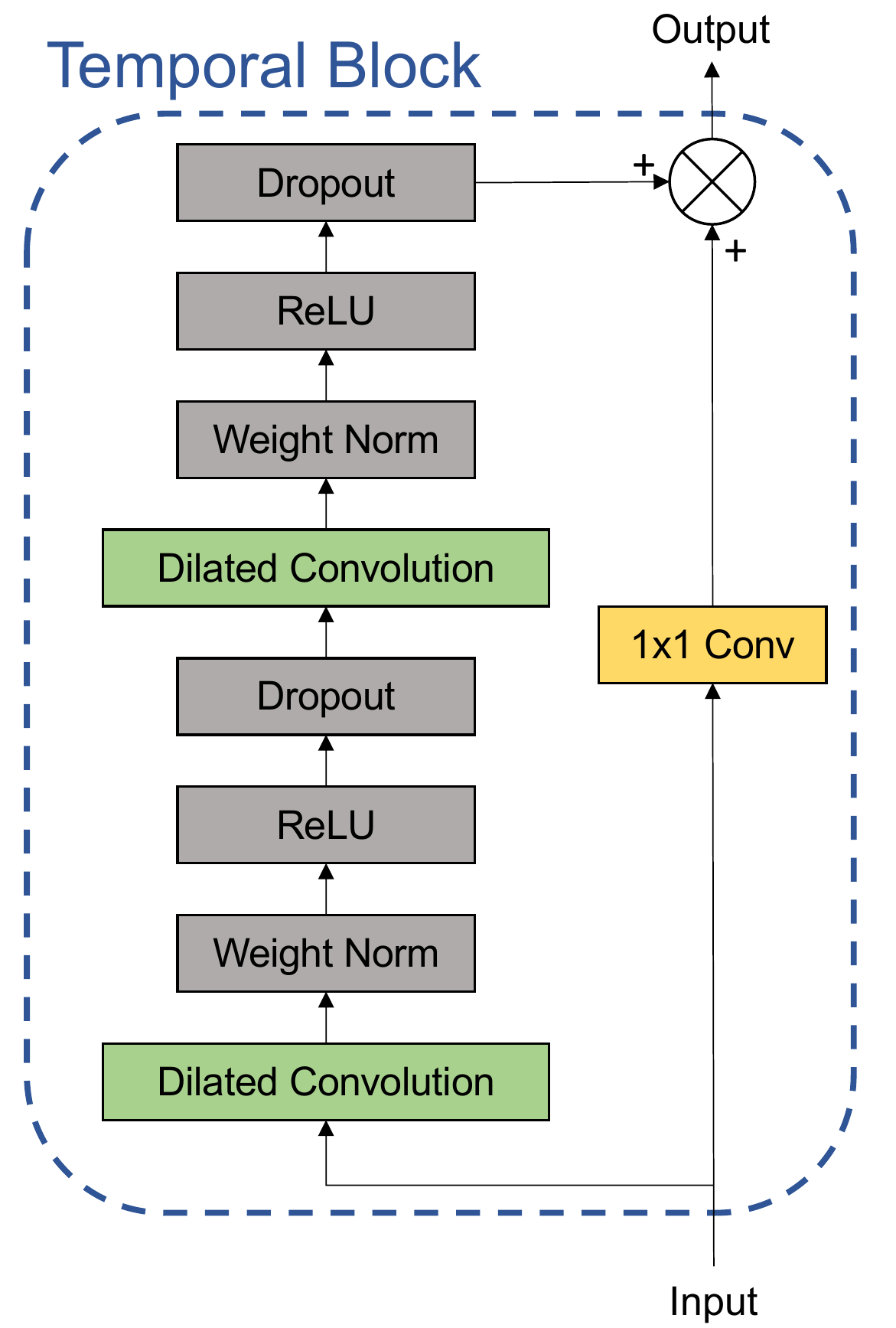}
  \caption{The structure of a Temporal Block.}
  \label{fig:tmpblk}
\end{figure}

TCN is based on a series of dilated 1-D convolutions organized into \emph{Temporal Blocks}. Each temporal block has the same basic structure. It has 2 convolution layers interspersed with weight normalization, dropout, and non-linearity layers. Figure \ref{fig:tmpblk} shows the organization of the various layers inside a temporal block.

The weight normalization layers reparameterize the weights of the network. Each weight parameter is split into 2 parameters, one specifying its weight, and the other its direction. This kind of reparameterization, as \cite[]{DBLP:journals/corr/SalimansK16} show, helps improve convergence. The Dropout layers randomly zero out layer outputs, which helps prevent overfitting. The ReLU nonlinearity layers allow the network to learn more powerful representations. Each Convolution layer adds padding to the input so that the output is of the same size as input. There is also a skip connection from the input to the output of each temporal block. A distinguishing feature of the TCNs is their use of dilated convolutions, that allows the network to have a large receptive field, i.e., how many samples of the input contribute to each output. The size of the dilation factor increases exponentially at each temporal block. With regular convolution layers, one would have to use a very deep network to ensure the network has a large receptive field. On the other hand, using sequential dilated convolutions allows the network to look at large parts of the input without having to use many layers. This enables TCNs to capture long term trends better than RNNs. Skip connections in the TCN architecture help stabilize training in case of deeper networks.
The concept of receptive field sits at the core of TCNs. Smaller convolution kernel sizes with fewer layers give the network a smaller receptive field, which allows it to capture local variations in sequential data well. However, such a network fails to capture the long term trends. On the other hand, larger kernel sizes with more layers give the network a large receptive field that makes it good at capturing long term trends, but not as good at preserving local variations. This is mainly due to the large number of successive convolutions which would dilute this information. This is also why adding skip connections to each residual block helps to overcome this drawback.

\section{Problem Formulation}
Let $\mathbf{X}=[\mathbf{x}^{1}, \mathbf{x}^{2}, ... \mathbf{x}^{N}]$ be a set of post-stack seismic traces where $\mathbf{x}^{i}$ is the $i^\text{th}$ trace, and $\mathbf{Y}=[\mathbf{y}^{1}, \mathbf{y}^{2}, ... \mathbf{y}^{N}]$ be the corresponding AI traces. A subset of X is inputted to the TCN in the forward propagation step. The network predicts the corresponding AI traces. The predicted AI traces are then compared to the true traces in the training dataset. The error between them is computed and is then used to compute the gradients. The gradients are then used to update the weights of the TCN in a step known as back-propagation. Repeated applications of forward propagation followed by backpropagation change the weights of the network to minimize the loss between the actual and predicted AI traces. We hypothesized that by treating both the stacked seismic trace $\textbf{x}^{n}$ and the corresponding AI trace $\textbf{y}^{n}$ as sequential data, we would be able to use the TCN architecture to learn the mapping $\mathcal{F}$ from seismic to AI. The training of the network can then be written mathematically as the following optimization problem:
\begin{equation}
    \hat{\Theta} = \underset{\Theta}{\text{argmin}}~ \frac{1}{N}\sum_{n=1}^{N}\mathcal{L}(\textbf{y}^{n}, 
    \mathcal{F}_{\Theta}(\textbf{x}^{n}))
    \label{eq:1}
\end{equation}
where $\mathcal{L}$ is a distance function between the actual and predicted AI traces, $\mathcal{F}$ represents the forward propagation of the TCN on the input seismic to generate the corresponding predicted AI trace, and $\Theta$ represents the network weights.

\section{Methodology}
\begin{figure*}[ht]
\centering
  \includegraphics[width=0.8\linewidth]{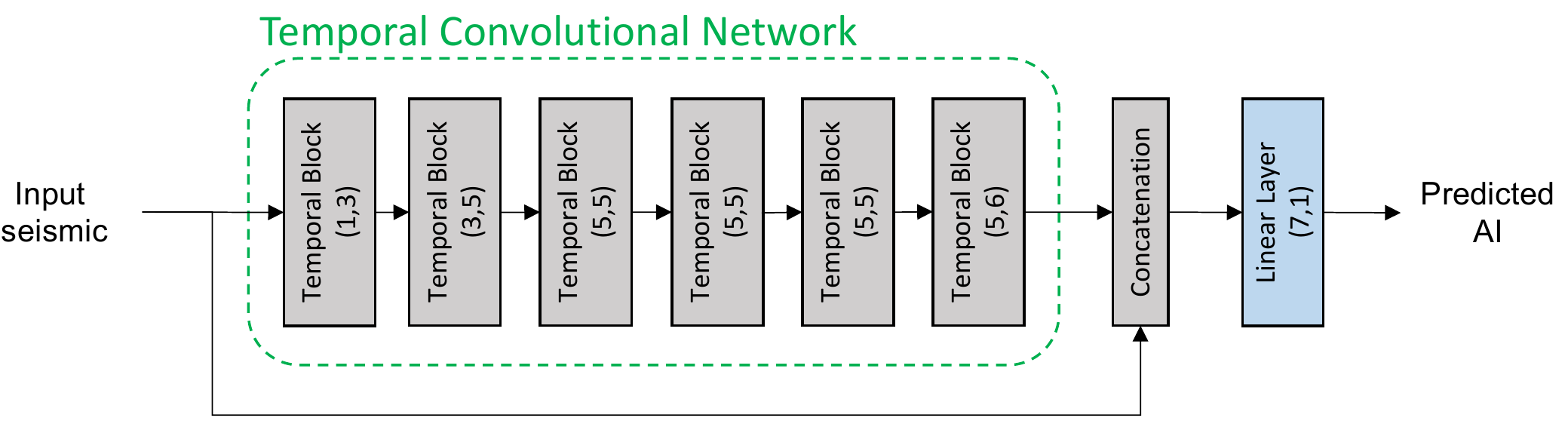}
  \caption{TCN architecture for predicting AI. The TCN consists of a series of 6 temporal blocks, the input and output channels for each specified in parentheses.}
  \label{fig:net}
\end{figure*}

The network architecture used is shown in Figure \ref{fig:net}. The seismic traces are passed through a series of temporal blocks. The output of the TCN is concatenated with the input seismic and then mapped to predicted AI using a linear layer. As discussed earlier, when using a larger kernel size with more layers, the network captures the low-frequency trend, but not the high-frequency fluctuations. On the other hand, with a smaller kernel size with fewer layers, the network captures the high frequencies but fails to capture the smoother trend. This is also why we concatenated the original seismic directly with the output of the TCN, so that any loss of high-frequency information due to successive convolutions in the temporal blocks might be compensated for. We found that this slightly improved the quality of our results. We experimented with different kernel sizes and number of layers, and found the numbers reported in Figure \ref{fig:net} worked best in terms of capturing both high- and low-frequency contents. 

\subsection{Training the network}
There is a total of 2721 seismic and corresponding AI traces from the Marmousi model over a total length of 17000m. We sampled both the seismic section and the model at intervals of 937m, to obtain a total of 19 training traces ($\leq 1\%$ of the total number of traces). We chose Mean Square Error (MSE) as the loss function. Adam was used as the optimizer with a learning rate of 0.001 and a weight decay of 0.0001. We used a dropout of 0.2, kernel size of 5, and 6 temporal blocks. The TCN internally also uses weight normalization to improve training and speed up convergence. We trained the network for 2941 epochs, which took about 5 minutes to train on a NVIDIA GTX 1050 GPU. Once the network had been trained, inference on the whole seismic section was fast and took only a fraction of a second.

\section{Results and Discussion}
\begin{figure}[ht]
    \centering
    \subfigure[Predicted AI]{
    \includegraphics[width=\linewidth]{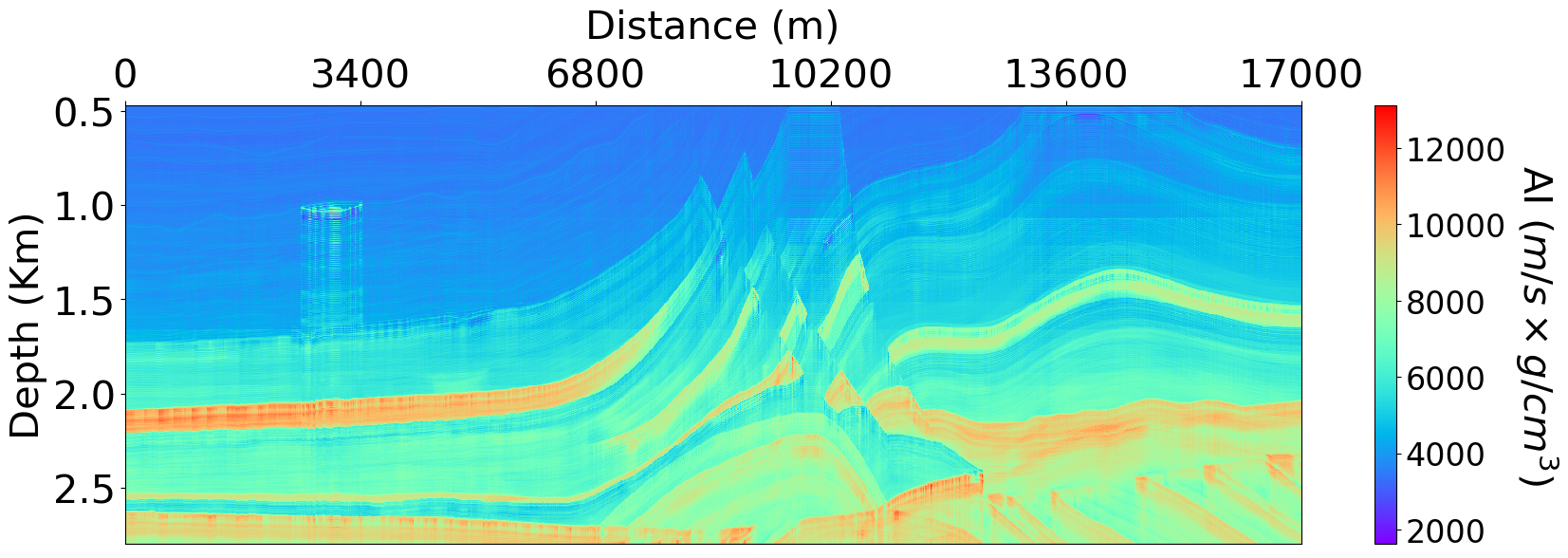}}
    
    \subfigure[True AI]{
    \includegraphics[width=\linewidth]{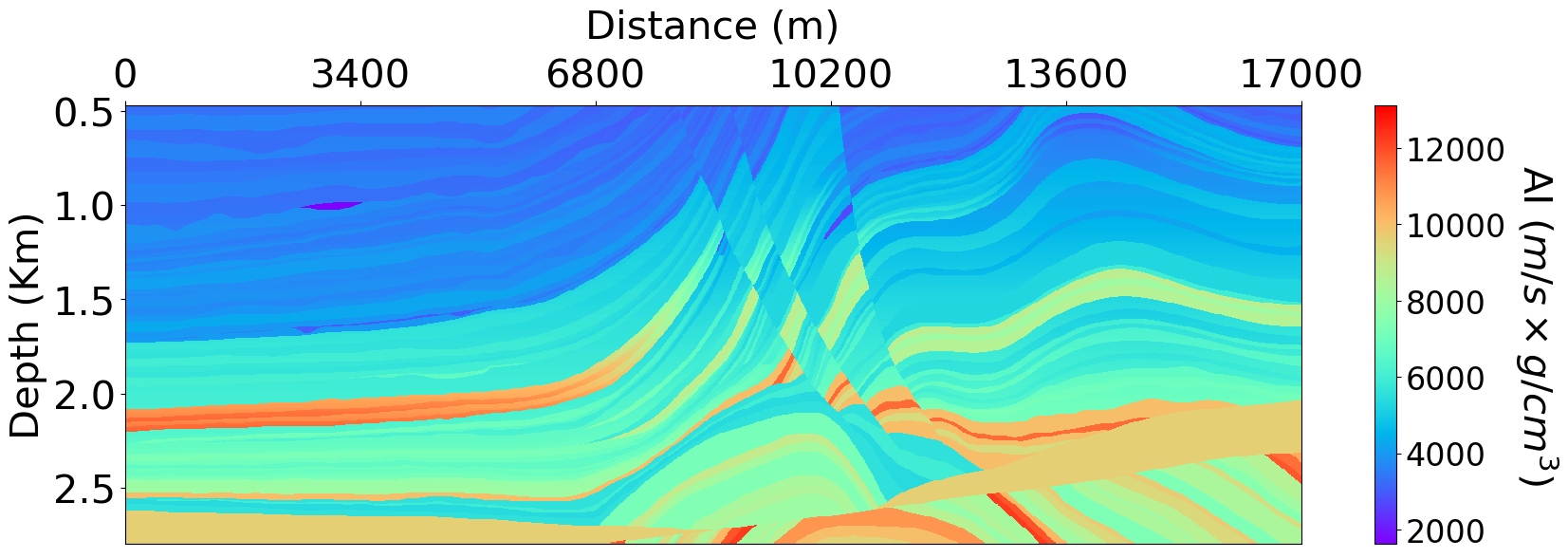}}

    \subfigure[Absolute Difference]{
    \includegraphics[width=\linewidth]{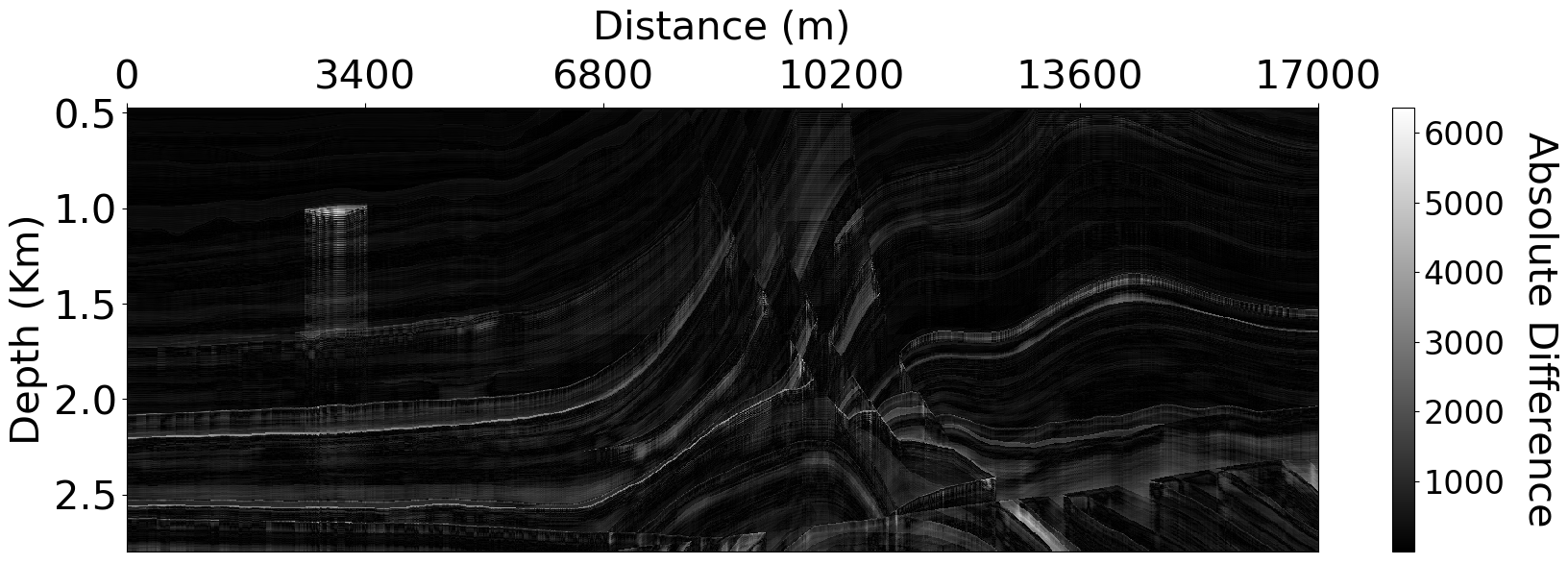}}
\caption{Comparison of the predicted and true Acoustic impedance sections of the Marmousi 2 model along with the absolute difference}
\label{fig:pred_sec}
\end{figure}

Figure \ref{fig:pred_sec} shows the predicted and actual AI, along with the absolute difference between the two. The predicted and actual AI sections show a high degree of visual similarity. The TCN is able to delineate most of the major structures. The difference image also shows that most of the discrepancy lies at the edge boundaries, which is because of sudden transitions in AI that the network is not accurately able to predict. 

%% Figure for image comes here

\begin{figure}[ht]
\centering
\includegraphics[width=\linewidth]{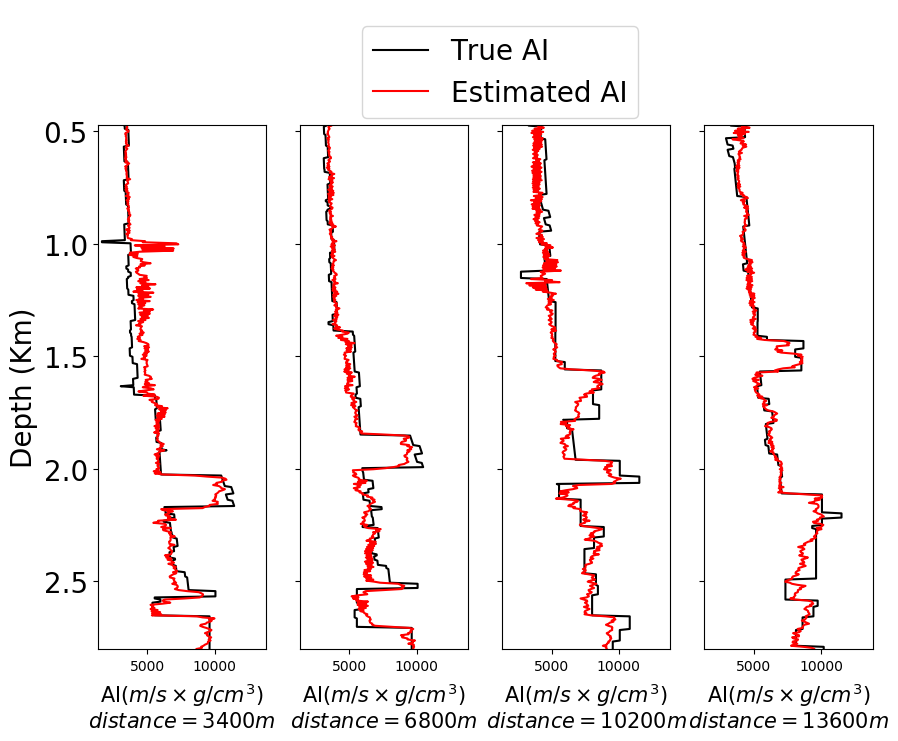}
\caption{Comparison of the predicted and true Acoustic impedance traces at selected locations along the horizontal axis.}
\label{fig:pre_trace}
\end{figure}

\begin{figure}[ht]
\centering
\includegraphics[width=0.7\linewidth]{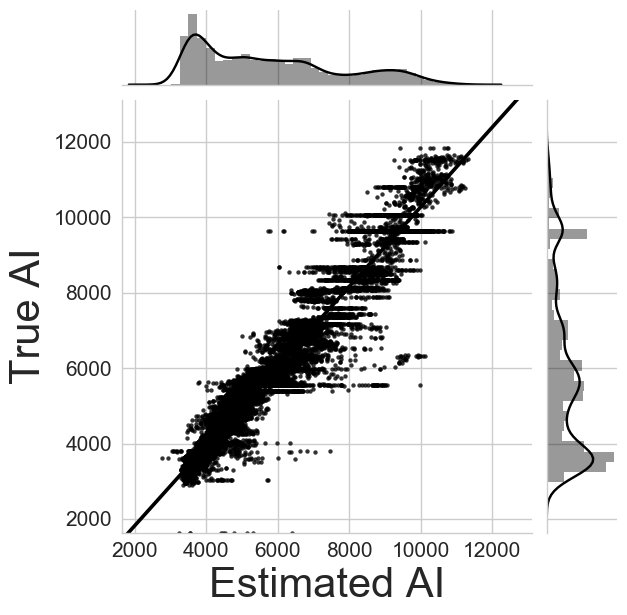}
\caption{Scatter Plot of the true and estimated AI}
\label{fig:scatter}
\end{figure}

We also show traces at 3400m, 6800m, 10200m, and 13600m, respectively in Figure \ref{fig:pre_trace}. As can be seen, the AI and estimated traces at each location agree with each other to a large extent. Figure \ref{fig:scatter} shows a scatter plot of the true and estimated AI. The scatter plot show that there is a strong linear correlation between the true and estimated AI sections.

For a quantitative evaluation of the results, we computed the Pearson's Correlation coefficient (PCC) and the coefficient of determination between estimated and true AI traces. PCC is a measure of the overall linear correlation between two traces. The coefficient of determination ($r^{2}$) is a measure of goodness of fit between two traces. The averaged values are shown in Table \ref{tab:results} for the training dataset and for the entire section (labeled as validation data). As can be seen, both the training and validation traces report a high value for the PCC and $r^{2}$ coefficients which confirms that the network was able to learn to predict AI from seismic traces well and to generalize beyond the training data.

%% Table begins
\begin{table}[ht]
\centering
 \begin{tabular}{|c|c|c|} 
 \hline
\textbf{ Metric} & \textbf{Training} & \textbf{Validation}  \\
 \hline
 PCC & 0.96 & 0.96 \\ 
 \hline
 $r^{2}$ & 0.91 & 0.91 \\
 \hline
%  Loss & 0.0838 & 0.0800  \\
%  \hline
\end{tabular}
\caption{Performance metrics for both the training and validation datasets.}
\label{tab:results}
\end{table}

\section{Conclusion}
In this work, we proposed a novel scheme of predicting acoustic impedance from seismic data using a Temporal Convolutional Network. The results were demonstrated on the Marmousi 2 model. The proposed workflow was trained on 19 training traces, and was then used to predict Acoustic Impedance for the entire Mamrousi model. Quantitative evaluation of the predicted AI (PCC $\approx 0.96$, and $r^2\approx0.91$) shows great promise of the proposed workflow for acoustic impedance prediction. Even though the proposed workflow has been used for AI estimation in this paper, it can be used to predict any other property as well. Indeed, Temporal Convolutional Networks can be adapted to any problem that requires mapping one sequence to another.

\section{Acknowledgements}
This work is supported by the Center for Energy and Geo Processing (CeGP) at Georgia Institute of Technology and King Fahd University of Petroleum and Minerals (KFUPM). 

\bibliographystyle{seg}  % style file is seg.bst
%\bibliography{references}  %  Comment out this command

\begin{thebibliography}{}
\itemsep0pt

\bibitem[Al-Anazi and Gates, 2012]{ALANAZI201264}
Al-Anazi, A., and I. Gates,  2012, Support vector regression to predict
  porosity and permeability: Effect of sample size: Computers \& Geosciences,
  {\bf 39}, 64 -- 76.

\bibitem[Alfarraj and AlRegib, 2018]{alfarraj2018petrophysical}
Alfarraj, M., and G. AlRegib,  2018, Petrophysical property estimation from
  seismic data using recurrent neural networks, {\it in} SEG Technical Program
  Expanded Abstracts 2018: Society of Exploration Geophysicists,  2141--2146.

\bibitem[{AlRegib} et~al., 2018]{regib}
{AlRegib}, G., M. {Deriche}, Z. {Long}, H. {Di}, Z. {Wang}, Y. {Alaudah}, M.~A.
  {Shafiq}, and M. {Alfarraj},  2018, Subsurface structure analysis using
  computational interpretation and learning: A visual signal processing
  perspective: IEEE Signal Processing Magazine, {\bf 35}, 82--98.

\bibitem[Bai et~al., 2018]{Bai}
Bai, S., J.~Z. Kolter, and V. Koltun,  2018, An empirical evaluation of generic
  convolutional and recurrent networks for sequence modeling: CoRR, {\bf
  abs/1803.01271}.

\bibitem[Biswas et~al., 2018]{biswas}
Biswas, R., A. Vassiliou, R. Stromberg, and M. Sen,  2018, Stacking velocity
  estimation using recurrent neural network: , 2241--2245.

\bibitem[Calder On-macas et~al., 1999]{article}
Calder On-macas, C., M. Sen, and P. Stoffa,  1999, Automatic nmo correction and
  velocity estimation by a feedforward neural network: GEOPHYSICS, {\bf 63}.

\bibitem[{Chaki} et~al., 2015]{Chaki}
{Chaki}, S., A. {Routray}, and W.~K. {Mohanty},  2015, A novel preprocessing
  scheme to improve the prediction of sand fraction from seismic attributes
  using neural networks: IEEE Journal of Selected Topics in Applied Earth
  Observations and Remote Sensing, {\bf 8}, 1808--1820.

\bibitem[Das et~al., 2018]{das2018}
Das, V., A. Pollack, U. Wollner, and T. Mukerji,  2018, Convolutional neural
  network for seismic impedance inversion, {\it in} SEG Technical Program
  Expanded Abstracts 2018:  2071--2075.

\bibitem[Dauphin et~al., 2016]{Dauphin}
Dauphin, Y.~N., A. Fan, M. Auli, and D. Grangier,  2016, Language modeling with
  gated convolutional networks: CoRR, {\bf abs/1612.08083}.

\bibitem[Johnson and Zhang, 2015]{johnson}
Johnson, R., and T. Zhang,  2015, Semi-supervised convolutional neural networks
  for text categorization via region embedding, {\it in} Advances in Neural
  Information Processing Systems 28: Curran Associates, Inc.,  919--927.

\bibitem[Kalchbrenner et~al., 2016]{Kalchbrenner}
Kalchbrenner, N., L. Espeholt, K. Simonyan, A. van~den Oord, A. Graves, and K.
  Kavukcuoglu,  2016, Neural machine translation in linear time: CoRR, {\bf
  abs/1610.10099}.

\bibitem[Lipari et~al., 2018]{GAN}
Lipari, V., F. Picetti, P. Bestagini, and S. Tubaro,  2018, A generative
  adversarial network for seismic imaging applications: Presented at the .

\bibitem[Salimans and Kingma, 2016]{DBLP:journals/corr/SalimansK16}
Salimans, T., and D.~P. Kingma,  2016, Weight normalization: {A} simple
  reparameterization to accelerate training of deep neural networks: CoRR, {\bf
  abs/1602.07868}.

\bibitem[van~den Oord et~al., 2016]{Oord}
van~den Oord, A., S. Dieleman, H. Zen, K. Simonyan, O. Vinyals, A. Graves, N.
  Kalchbrenner, A.~W. Senior, and K. Kavukcuoglu,  2016, Wavenet: {A}
  generative model for raw audio: CoRR, {\bf abs/1609.03499}.

\end{thebibliography}

%%% Paste the contents of the bbl file here %%%%

%%%%%%%%%%%%%%%%%%%%%%%%%%%%%%%%%%%%%%%%%%%%%%%%

\end{document}